# Use of frit-disc crucible sets to make solution growth more quantitative and versatile.

P. C. Canfield[1,2] and T. J. Slade[1]

1 Ames National Laboratory, Iowa State University, Ames Iowa 50011

2 Department of Physics and Astronomy, Iowa State University, Ames Iowa, 50011

**Abstract**

The recent availability of step-edge, frit-disc crucible sets (generally sold as Canfield Crucible Sets or CCS) has led to multiple innovations associated with our group's use of solution growth.  Use of CCS allows for the clean separation of liquid from solid phases during the growth process.  This clean separation enables the reuse of the decanted liquid, either allowing for simple, economic, savings associated with recycling expensive precursor elements or allowing for the fractionation of a growth into multiple, small steps, revealing the progression of multiple solidifications.  Clean separation of liquid from solid phases also allows for the determination of the liquidus line (or surface) and the creation, or correction, of composition-temperature phase diagrams.  The reuse of clean decanted liquid has also allowed us to prepare liquids ideally suited for the growth of large single crystals of specific phases by tuning the composition of the melt to the optimal composition for growth of the desired phase, often with reduced nucleation sites.  Finally, we discuss how solution growth and CCS use can be harnessed to provide a plethora of composition-temperature data points defining liquidus lines or surfaces with differing degrees of precision to either test or anchor artificial intelligence and/or machine learning based attempts to augment and extend the limited experimentally determined data base.

**Dedication (by PCC)**

There must have been something in the air or water in Ames, Iowa. (Maybe there still is.)  Many of the researchers in Condensed Matter Physics and Solid State Chemistry associated with Ames Laboratory and Iowa State University were productive, engaged, collaborative and influential in the fields of new materials discovery, growth and characterization, making the hybrid Ames Lab / Iowa State University research environment a fantastic place to raise a research group.  Gordon Miller (Gordie) is one of these researchers.  He has been endlessly generous with his time, his interest, his assistance, and his insight.  He has also been tolerant (like a kindly gent with an over-enthusiastic puppy) with stray physicists asking repetitive and clearly naïve questions.  Gordie has been one of my "go-to-chemists" for questions about bonding/banding, widths of formation, X-ray interpretation, and advice for about how to navigate the realm that exists between chemistry, metallurgy and physics.   It is a pleasure to contribute a paper to a collection in honor of Gordie's career; we hope it both entertains and enlightens.

**Introduction**

Solution growth of single crystals works best when the growth starts in a high temperature, single phase liquid region of the composition-temperature phase diagram with a composition that intersects a liquidus line or surface for primary solidification upon appropriately slow cooling. Whereas the sentence above is correct, compact, perhaps even well written, it contains several examples of, "the devil is in the details". Some of these details are: determining that the melt is in a homogeneous single phase at high temperature, making sure it is free of known, or unknown, undissolved materials, making sure it intersects a liquidus line/surface, and determining at what temperature this takes place, just to name a few. In this paper we hope to show how the use of frit-disk crucible sets allows for systematic and quantitative methods to address many of these "devils". We will build on a series of papers that describe many of the basics[1-3] as well as specifics[4-6] of solution growth and focus on how the widening use of frit-disc crucible sets, commonly sold as Canfield Crucible Sets (CCS)[7,8] is markedly improving how our lab, and others, is approaching new materials discovery and crystal growth.

We will outline how use of CCS allows for simple and clean reuse of decanted liquid, how they can be used for determination of parts of temperature – composition phase diagrams, and how they can be used to prepare clean melts and create optimal single phase solutions for subsequent crystal growth. Our intent is to provide detailed examples that can be readily generalized to other, new systems. We also hope that with the quantification that comes with CCS use will allow for more systematic collection and creation of temperature-composition data points for liquidus lines, or surfaces, of primary solidification for binary, ternary, quaternary and higher phase diagrams. Given the increased potential (or at least promise) of artificial intelligence / machine leaning (AI/ML) for phase diagram refinement, the importance or readily available experimental data points that can be used to test or anchor multinary phase diagrams should not be understated, but often is.

**Canfield Crucible Sets**

It is useful to begin with a brief outline of the basics of solution/flux growth of single crystals[1,2]. Very broadly, one starts by weighing precursors (elements or compounds) into a crucible or ampule, sealing the materials under some form of protective atmosphere or vacuum, and heating to a suitably high temperature such that all of the starting materials have melted into a homogenous, single phase, liquid. The melt is then slowly cooled to some target temperature, during which time crystal growth (hopefully) occurs. Finally, the excess liquid phase is rapidly separated from the products by pouring off the excess liquid, a process generally known as decanting. In some cases, motivated by fundamental need or lack of technology, instead of decanting chemical etchants or mechanical removal can alternatively, be used. If one is working with a well understood system, the above procedure should reliably produce crystals of the desired phase, and this would be the end of the story. Of course, such well understood chemical systems are rare (most often only binary combinations), and information on ternary (or higher multinary) phase diagrams is at best sparse, making trial-and-error usually necessary to determine what, if any, conditions are optimal to grow a desired material (or discover new phases). In practice, these trials will generically result in any of the following: 1) a single, crystalline phase separated from a decant; 2) a mixture of crystalline phases separated from the decant; 3) no crystalized phases and all material decanted; or 4) no decanted liquid (the desired phases may have formed, but are trapped in the solidified matrix of the un-decanted, and now solidified, excess liquid). In each of these cases, it is often extremely useful to be able to re-use portions of the initial growth attempt, i.e. to re-melt the decanted

(or un-decanted) liquid and bring to a yet lower (or higher) temperature. To accomplish this in a controlled manner, it is necessary to have a means of capturing both the products and any decanted liquid, without significant loss or contamination of material. We outline below how Canfield Crucible Sets (CCS) allow us to accomplish this in a routine manner, providing history and examples from our prior research.

The CCS was born to the usual parents: need and cost. Whereas use of filtration materials has been part of solution growth for many decades[2], with broken silica, fiberfrax or silica wool commonly used, these materials contaminated the decanted liquid, rendering further use or even analysis of it complex at best and virtually impossible at worst. When we were contending with reactive melts, such as those containing significant percentages of alkali, alkali earth, or rare earth elements, we implemented the use of what we called three-cap Ta-tube growths[1, 3, 6]. For these growths we used a Ta-tube with three press-fit Ta caps. The upper and lower ones are welded in place with the middle one having small holes drilled into it, to act as a filter. These 3-cap Ta-tubes did allow us to potentially reuse decant, but they posed multiple problems ranging from the need to cut them open to retrieve the products and decant (rendering them non-reusable) to being expensive and time consuming to make. This said, they do allow for some of the advantages of CCS, as will be exemplified below when $La_2Ni_7$ and the La-Ni binary phase diagram are discussed.[9]

Based on our experience with the clean and reusable decants that we could get from 3-cap-growths, we used BN rods and machined rather thick walled crucible sets that consisted of two ~3 ml volume cylindrical crucibles with internally threaded tops and a nut-like intermediate piece that has multiple holes drilled in a central plate and threads on both sides so that it could screw into both crucibles.[10] This set allowed us to use (and reuse) Pt-B based melts that we used in our (failed) attempts to grow $MgB_2$ single crystals at ambient pressure. Although this BN threaded crucible set did not lead to published work, it was a key evolutionary step.

The next step toward the CCS was to have 2- and 5-ml crucible sets made from alumina with a rather finely threaded frit disc that screwed into threads on the inner surfaces of two cylindrical crucibles. A squat, version of one of these is shown in Figures 1, 4, and 5 of ref. 10. Whereas these were exceptionally beautiful objects, they were quite expensive to have made and, in addition, were not always reusable given that the rather fine threads often got glued shut by infiltration of decanted liquid . This meant that in order to open the closed set at least one crucible and often both crucibles had to be broken open. This said, they offered very clean separation of the grown crystals (the solid phase) from the remaining liquid.

The key idea that allowed the CCS to be readily fabricated and used was the realization that, instead of a thread, a step-edged-frit disk would work well[7]. As can be seen in Fig. 1 the alumina frit is not a simple, uniformly thick disk, but rather has an outer diameter, in the middle, that matches the outer diameter of the crucibles and an offset upper and lower outer diameter that is a little smaller than the inner diameter of the crucibles. This design allows the step-edge frit-disk to self-center into the open ends of the two crucibles and is much easier to manufacture. This design also avoids much of the problem of the threaded frit crucibles getting glued shut, but it does allow some spillage if the frit gets plugged by plate-like or polycrystalline growths. If spillage is a problem, then a modification of the CCS can be used (as will be discussed at the end of this paper).

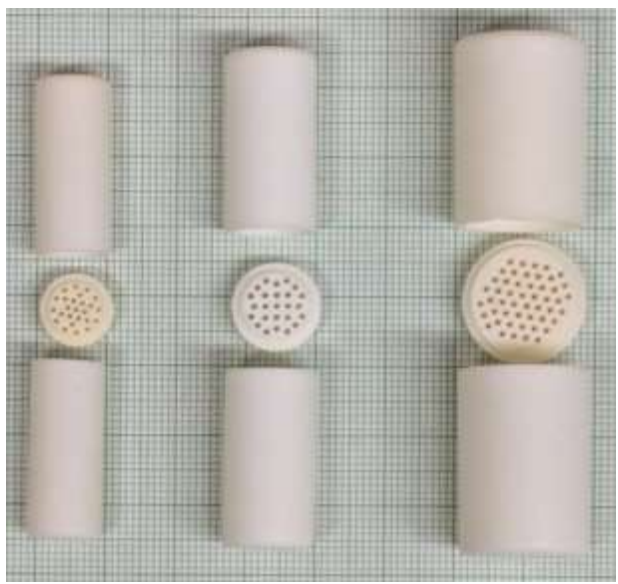

Figure 1: (left to right) 1.8 ml, 2 ml, and 5 ml $Al_2O_3$ CCS on a mm grid. In typical reactions, the bottom "growth" side holds the precursor materials, the frit-disc is used to separate the solid phase(s) from excess liquid/flux when decanting, and the decanted liquid is cleanly captured in the top "catch" crucible.

Returning to this idea of "need and cost". We should note, that before adopting use of CCS as our default growth crucible, we used pairs of 2 ml crucibles sold by certain beer company in which the upper, catch crucible, was filled with silica wool.[1-3] These crucibles had been used in our group for over 20 years and, historically, we had been able to get non-trivial bulk discounts for large orders. At much the same time we had developed the threaded crucibles, the beer company dramatically increased their prices and did away with any bulk discounts. The initial design goal for the CCS was to have a three-piece *set* cost less than a single crucible of the same size from the beer company. This was achieved and, checking websites recently, it is still the case. At this point CCS are available in sizes ranging from 1.7 ml up to 10 ml and have been fabricated out of alumina, zirconia and boron nitride.[8] For our group, changing to default CCS use led to a rare example of both better and cheaper procedures.

**Re-use of decanted liquid**

The frit-disk, at the most basic level, allows for the clean separation of solid phase (if any) from the remaining liquid. If we are more precise and spell out all the assumptions and examine all the warts and wrinkles, then we can say that by decanting through the alumina frit-disk we are not making the liquid any dirtier than it was in the alumina crucible and that we have separated out solids (or conglomerates of solids) larger than the holes in the frit. If there are tiny grains of solid suspended in the liquid that are not agglomerating, then these may pass through the frit. All of this said, the use of the alumina frit is vastly superior to decanting into a plug of silica wool or fiberfrax[2] since these fibrous filters hopelessly contaminate the decanted liquid. With this clean separation of the liquid and solid phases at a given decanting temperature, use of the CCS allows for relatively quick and inexpensive tests of the melt as well as recycling/reuse of melts that crystals have been grown out of.

For example, one of the premises of solution growth, at least reproducible solution growth, is that, at the highest temperature, the melt is in a single-phase liquid part of the composition-temperature phase diagram.  Sometimes, it is important to make sure that this is indeed the case, especially if the elements have a mixture of properties, such as refractory (transition metal) and volatile (pnictogens, chalcogenides, etc.) or if details of the ternary or higher phase diagram are unknown.  In these cases, we can heat the initial composition of elements to the highest temperature we plan to use, sit there for what we hope is enough time to reach equilibrium, and then decant.  Given that we use amorphous silica tubes to hold the CCS growths, our maximum temperature is often 1200°C.  If we get a total decant at the growth's highest temperature, i.e. the full mass of our initial melt passes through the frit, then we can feel moderately secure in thinking we have a single-phase liquid.  We can then reuse the decanted liquid for subsequent growth, secure in our assumption that we are indeed starting from a single-phase liquid.  In addition, if we are working with a volatile element we can see if there is any weight loss, often correlated with some degree of condensation of the volatile on the inner surface of the silica.

If there is not a full decant, if we find solid phase on the growth side of the CCS rather than all the decant passing through to the catch side, then we know that the initial stoichiometry was not in a single phase region of the composition phase diagram at the temperature of the decant (or that there were dynamics associated with melting that were not completed in the time we sat at the highest temperature).  When this is the case, we can recombine the solid from the growth side with the decanted liquid and either heat to a higher temperature (if we were not at 1200°C) or try a longer, high-temperature soaking period.  If neither of these corrective measures produce a single-phase liquid, then we can proceed with the decanted liquid as our starting melt and engage in fractionation, as will be described in the next section, below.

In recent years we have been exploring and developing a number of high temperature solutions based on transition metal rich eutectics[11-18].  Before we start adding ternary or even quaternary elements to the melt, it is useful to confirm that the published binary phase diagram is correct at the level of detail needed and that the elements can be safely and reliable combined to make the melt.  We can take the Rh-S binary phase diagram as an example.  Whereas the purported composition of the most Rh-rich eutectic is $Rh_{0.65}S_{0.35}$, when we tried growth from this composition we ended up with significant growth of elemental Rh.  This, of course, implied that the eutectic composition was more S-rich.  We found that the eutectic composition is closer to $Rh_{0.60}S_{0.40}$.[17]  Not only have we been able to use this eutectic region as a starting point for the discovery of new, ternary Rh-S-X superconductors[11, 12], but we also have been able to grow single crystals of the mineral known as Miassite[17, 18], a rare example of a mineral that is superconducting in synthetic form.

For many of the refractory – volatile systems we have worked with in recent years, if we can create a single phase liquid and cool a eutectic composition to a little above the eutectic temperature and still have a complete decant with no apparent crucible attack or vapor pressure loss, then we feel that we "own the eutectic" and proceed to use binary compositions around the eutectic composition as high temperature solutions for the growth of binary, ternary and quaternary compounds.  Examples of eutectics we considered "owned" are:  Co-S, Mn-P, Pt-P, Pd-P, Rh-S, Pd-S, Pd-Se, Pt-Te, and others.

In some cases, compounds of interest need to be grown out of solutions rich in very expensive components.  When this is the case, *if* the material that crystallizes is single phase, with a well-defined stoichiometry and is well separated from the decanted liquid (meaning there is minimal liquid left on the crystallized phase), *then* you can infer the stoichiometry of the decanted liquid by subtracting off the crystallized material from the composition of the original charge.  With this knowledge, you can then determine the composition of the decant and reuse it.

For example, when we wanted to grow single grains of icosahedral R-Cd quasicrystals for neutron scattering studies we needed to use isotopically enriched $^{114}$Cd.  $^{114}$Cd is a less-neutron absorbing isotope that is purchased by the mg rather than gram.  Given that the growth of R-Cd is out of a vast excess of Cd (as shown for the case of Gd-Cd in Figure 2b below), simply discarding the decanted solution would be an exceptionally expensive proposition.  Instead we cooled a melt with initial composition of $Tb_{0.008}Cd_{0.992}$ from the single-phase liquid region down to 330 C, decanted, weighed the i-Tb$^{114}$Cd single grains, determined the composition of the decanted material, added more Tb and repeated growth.  After doing this several times, the remaining decant was used to grow single crystals of Tb$^{114}$Cd$_6$ (a crystalline approximant of the i-Tb$^{114}$Cd QC) for comparison studies.[19-22]  In this manner we were able to much more efficiently use the $^{114}$Cd. We have employed similar strategies when growing Rh-, Pt-, Pd- and Au-rich materials.

**Determination of the liquidus line or surface**

Given that we can infer the composition of the decanted solution by measuring either the mass of the decant and/or the mass of the grown, single phase sample, we can put a point on the liquidus line (or surface) at the decanting temperature.  To be very precise about the assumptions here, the accuracy of the data point, in terms of composition, depends on how well we can measure the mass of the clean single phase solid that is left on the growth side or infer this mass based on the mass of the decant.  If there are relatively few crystals with very little remaining decant on them, then we can infer the composition of the remaining melt quite well.  There is relatively small uncertainty in the temperature value of this data point given that, even at highest temperatures we can decant within 10 degrees of a eutectic solidification; the decanting process is quick, less than 6 seconds from furnace to decanted state.

An example of how this can allow us to determine the liquidus line in a binary phase diagram is shown in Fig. 2b for the Gd-Cd binary phase diagram.   GdCd$_6$ grows as large, well faceted crystals that cleanly separate from the liquid phase during decanting, giving mirrored surfaces.[22]  This allows us to determine four points on the liquidus line for GdCd$_6$, each point resulting from a growth that had an initial composition with higher Gd content than the remaining decant.[19, 22]  For the data points on the i-GdCd liquidus, we used starting compositions that intersected the liquidus surface below the peritectic decomposition of the QC phase, i.e. that yielded single phase growth of pentagonal dodecahedral single grains.  As shown in the insets of Figs. 2a and 2b, these QC-grains have clean facets and allow for inference of the composition of the remaining liquid phase.

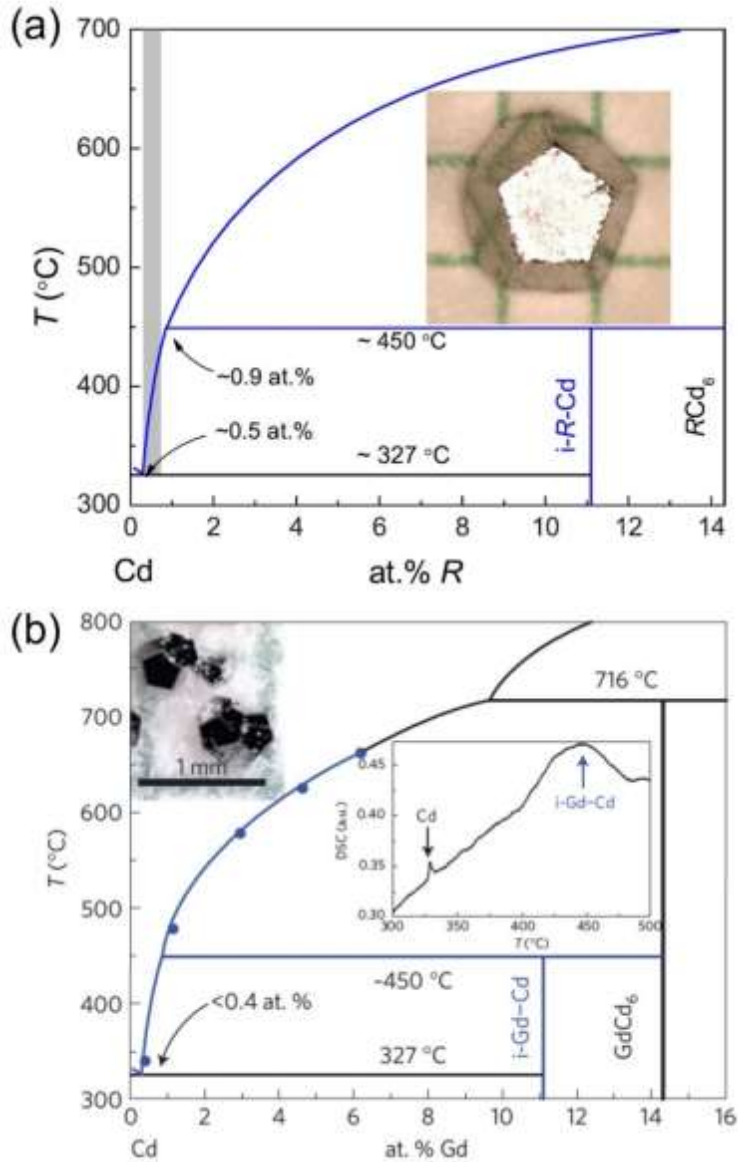

**Figure 2:** (a) General R-Cd binary phase diagram for Cd-rich compositions[20]. The grey shaded area shows the compositions that intersect the liquidus line for i-R-Gd quasicrystal growth, and the inset shows a single i-Gd-Cd quasicrystal on a millimeter grid. (b) Binary phase diagram for Gd-Cd [19]. The points on the liquidus line were determined by decanting growths at different temperatures and weighing the crystalline (GdCd$_6$) or i-Gd–Cd product (see text for more details). The central inset in (b) shows differential scanning calorimetry data from single grains of the quasicrystalline phase taken on warming through their thermal decomposition at roughly 450 °C. The small feature near 320 °C is the melting of the small amount of residual Cd flux on the surfaces of the quasicrystal grains. The upper inset is a picture of several small i-Gd-Cd quasicrystals.

Whereas phase diagram determination is relatively simple with an alumina CCS, we have been able to perform similar refinements with Ta-3-cap crucibles as well[1, 3]. A recent example of this is our work on the La-Ni binary phase diagram associated with our growth of large single crystals of $La_2Ni_7$.[9, 23-25] In our initial attempts to grow $La_2Ni_7$, we used the most recent La-Ni binary phase diagram in the ASM online database[9, 26, 27] which has the exposed liquidus line for $La_2Ni_7$ existing between 63% at. Ni at 979 °C and 57% at. Ni and 802 °C. When we cooled a melt of $La_{38}Ni_{62}$ from 1050 to 820 °C and decanted[9], we found that the melt was still in a single-phase liquid state (i.e. no products were formed and all the liquid passed through the filter into the "catch" side of the Ta crucible). When we re-melted the material, slow cooled, and then decanted again at 750 °C, we found a mixture of solid $LaNi_3$ and $La_7Ni_{16}$ (in roughly a 7:3 ratio as suggested by powder x-ray diffraction) in addition to decanted liquid. These results are inconsistent with Refs. 26 and 27 and suggest that the liquidus line for $La_2Ni_7$ is shifted to higher Ni concentrations. An earlier assessment of the La-Ni binary phase diagram[28] places the liquidus line for $La_2Ni_7$ between ~68% at. Ni at 976 °C and ~65% at. Ni at 811 °C. When we performed growth using a starting composition of $La_{33}Ni_{67}$ and cooled from 1020 to 820 °C, we produced single-phase $La_2Ni_7$ single crystals, allowing for an evaluation of the decanted liquid composition, which was ~65% at. Ni. The fact that we needed to use sealed Ta, fritted crucibles made this a more painstaking process than it would have been if we could have used alumina CCS, but the reactivity of the melt precluded that option.

Another mode of phase diagram determination is exemplified by our recent use of CCS for fractionation studies. In this case we start with a given initial composition of a multinary melt and through successive decanting and reuse of decanted liquid we can determine at what temperature ranges we have primary solidification of a cascade of multiple phases. This is well illustrated by recent work we have done on $RPd_3S_4$ compounds[5, 29, 30]. When we first grew single crystalline $CePd_3S_4$ we chose an initial composition by considering the beauty of the deep eutectic on the Pd-rich side of the Pd-S binary phase diagram and added a small amount of Ce to it.[4, 5] Starting with this $Ce_5Pd_{58}S_{37}$ initial composition, we ended up with mixed phase growth, containing binary Ce- and Pd- sulphides and small, sub-mm crystals of $CePd_3S_4$. In order to understand over what temperature ranges and in what order these phases were grown, we engaged in a series of relatively quick growths in which, for the (N+1)th growth we reused the decanted liquid from the preceding Nth growth. The results of such a fractionation study are shown in Fig. 3 for an initial melt composition of $Ce_5Pd_{58}S_{37}$.

Here, we first slowly heated the $Ce_5Pd_{58}S_{37}$ mixture to 1150°C, held at 1150°C for 8 h, after which the reaction ampule was removed from the furnace (at 1150°C) and the liquid fraction decanted. This temperature profile and the resultant material left in the growth side are shown in Fig. 3a. Here, all solids are captured in the initial "growth" crucible by the fritted-disc, while the liquid phase passes through the frit-disc and is caught in the second "catch" crucible. Clean capture of the liquid in the catch crucible without external contamination (i.e. from silica wool), makes it possible to reuse the decanted mixture for subsequent experiments. We found that after decanting at 1150°C, $Ce_2S_3$ was present in the initial growth crucible, clearly indicating that some $Ce_2S_3$ did not enter the liquid phase and persisted (forming a framework, along with small amounts of oxi-sulfides, for subsequent nucleation of multiple phases in the initial growths[4, 5]). In order to ensure that we started with a single-phase liquid, we used this first step as a way of preparing our initial high temperature liquid.

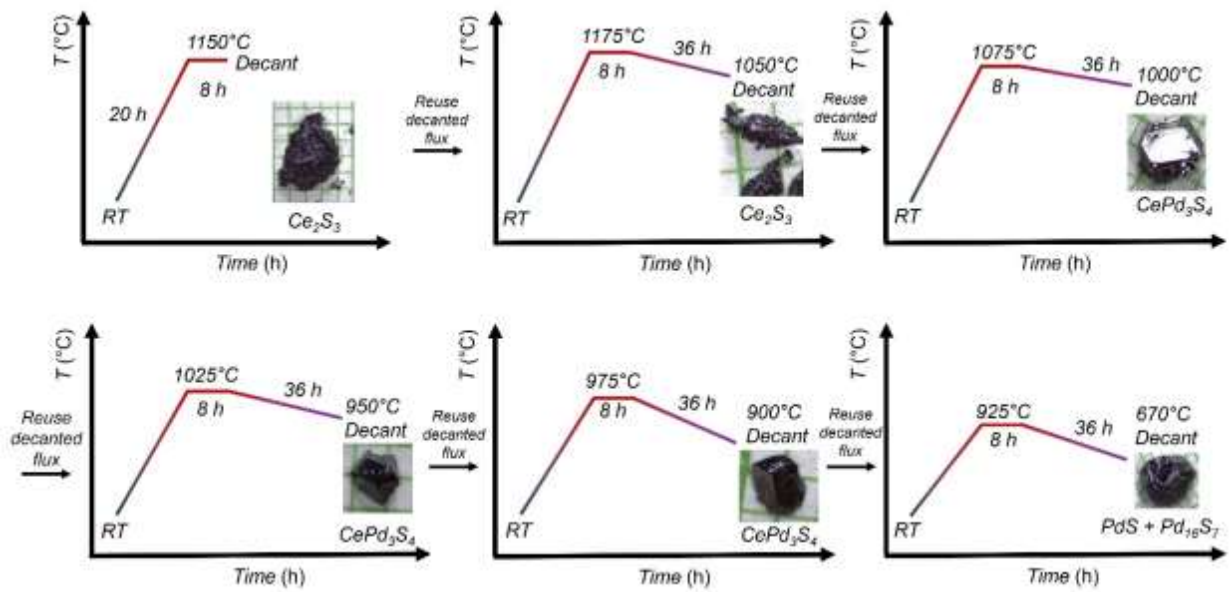

**Figure 3:** Schematic depicting the fractionation process using CCS. The starting materials are initially heated to a high temperature (normally just below 1200°C if silica tubes are used) and after dwelling, the liquid phase is decanted and captured in the "catch" side of the CCS, allowing for reuse in subsequent reactions. By iteratively re-melting and decanting at decreasing temperatures, the different solid phases, and approximate temperatures over which they crystallize, for any given starting combination, can be monitored. Note that each step heats up to an initial temperature 25°C above the decanting temperature from the previous step to ensure complete melting of all phases. The figure depicts the results for a growth with initial composition $Ce_5Pd_{58}S_{37}$, showing that $Ce_2S_3$ forms above ~1050°C, $CePd_3S_4$ between ~1050-900°C, and PdS and $Pd_{16}S_7$ below ~670°C.

After the first step, we next assembled a new crucible set, using the original catch crucible containing the decanted material as the new growth side. We heated this second growth to 1175°C (25 °C above the first decanting temperature to ensure that all material melts), held for 8 h, cooled the furnace to 1050°C in 36 h, and again decanted the remaining liquid (see Fig. 3b). We repeated this process of using the solidified liquid captured by the catch crucible during decanting, which was remelted and cooled to decanting temperatures in 50°C increments between 1150 – 670 °C. By carefully analyzing the products contained in the growth crucible after each step, we determined that the $Ce_2S_3$ primarily forms above 1050°C, the $CePd_3S_4$ crystallizes between ~1050°C and 900 °C, and the Pd-S binaries below 700°C (decanting at 670°C gave a mixture of primarily PdS with a smaller amount of $Pd_{16}S_7$ and $Pd_4S$). A step-by-step breakdown of the fractionation process, with pictures of the products isolated after each step, is shown in Fig. 3.

Based on the results of our fractionation, we sampled the trajectory of our initial composition as it cooled and intersected liquidus surfaces and, as a result, we devised an optimized two-step process to grow large crystals of $CePd_3S_4$. First, a nominal composition of $Ce_5Pd_{58}S_{37}$ was loaded into an alumina CCS and sealed in a fused silica tube as described above. The tube was heated to 1150°C, held for 8 h, and cooled over 36 h to 1050°C, after which the liquid was decanted. The tube was opened, all $Ce_2S_3$ and oxisulphides discarded, and the captured decanted liquid reused in a new crucible set. The second crucible set was again warmed to 1075°C. After holding for 8 h, the furnace was slowly cooled over ~150 h to 900°C, at which point the remaining solution was decanted. After cooling to room temperature, the CCS was opened to reveal large, mirror-faceted crystals, an example of which is shown in Fig. 4. With these crystals we were able to study the evolution of ferromagnetic and orbital ordering with pressure in $CePd_3S_4$,[30] as well as discover and characterize charge order transitions and valence changes under pressure associated with the mixed valence $Eu^{2+}/Eu^{3+}$ atoms in $EuPd_3S_4$.[29] We should note that, theoretically, if we could determine the precise amount and composition of the high temperature sulphides and oxisulphides we discarded in the first step, then we could perform this growth in a single step with a corrected initial composition. In practice, though, this is difficult and, in addition, getting rid of possible oxide based nucleation sites with a first decant is good enhancing the chances for a small number of larger crystals growing.

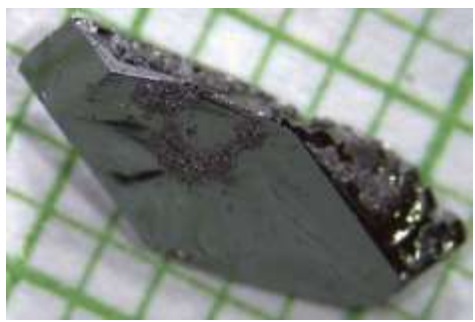

**Figure 4:** An image of a large single crystal of $CePd_3S_4$ grown using the optimized two-step process outlined in the text. Note that it is roughly 100 times larger than the crystal grown in our pre-fractionation growth attempts.

**Creating solution for growth**

Another use of CCS is preparing, or optimizing, solutions for growth of crystals. One important way of doing this is to make sure that single crystal growth starts from a single-phase-liquid part of the phase diagram. If the growth starts from a mixed phase state (i.e. some liquid, various solids), then nucleation is likely to occur at many, ill-defined, sites rather than at a relatively small number of sites. This generally means that the resulting crystals will be small, ill-formed and often contain second phase impurities. When we discussed fractionation in the section above, the first steps were to remove undissolved binary sulfates as well as any oxide-based slags (likely from slightly oxidized surfaces of the initial elemental rare earths) from the melt (Fig. 3a). By doing this we could allow the nucleation of single phase $CePd_3S_4$ in relatively few locations, leading to the growth of 10- to 100-times larger volume crystals.

Perhaps a more subtle, but equally important, example of preparing a solution can be seen in our recent growth of $Re_3Ge_7$ single crystals. As can be seen in Fig. 5, the reported liquidus line for growth of $Re_3Ge_7$ is shown as a (virtually) vertical line. For solution growth, this is exceptionally inauspicious given that getting the precise composition needed to intersect this line between 1132°C and 930°C is difficult at best and improbable at worst.  In the past this would be a repetitive, hit or miss effort.  With the CCS, and a little calm thought, it becomes clear that we can grow $Re_3Ge_7$ in a manner that is very reminiscent to how water soluble salt crystals have been grown for centuries[31]; we can create a perfectly saturated solution at 1132°C and then cool.  As a bonus, we go through an intermediate step that can remove any slag that might be involved.

We do this by first putting way too much Re into the initial melt. We put in $Re_{0.06}Ge_{0.94}$ into a CCS with the Re in the form of a single lump (way too big to fit through a frit hole).  We then heated to 1200°C, dwelled there for 12 hours, and cooled to 1130°C over 12 hours.  At that point we decanted the CCS, finding much of the lump Re still on the growth side and perfectly saturated Re-Ge solution (now solidified) on the catch side.  As a second step we took the decanted liquid, using it in a growth side of a CCS, heated to 1150°C, dwelled at 1150°C for 10 hours, cooled to 1135°C over 2 hours, and the slowly cooled to 940°C after which we decanted.  The result was reproducibly being able to grow rods of $Re_3Ge_7$. A picture of typical $Re_3Ge_7$ rods is shown in the inset to Figure 5.  With these rods we were able to study how pressure could suppress a metal-to-insulator transition and ultimately reveal a superconducting state.[32] The physics aside, $Re_3Ge_7$ offers a reminder that we can take advantage of reusing decanted liquid to prepare a single-phase liquid that is positioned precisely where we choose on a liquidus line or surface.  For growths with steep liquidus lines, this can be a great advantage.

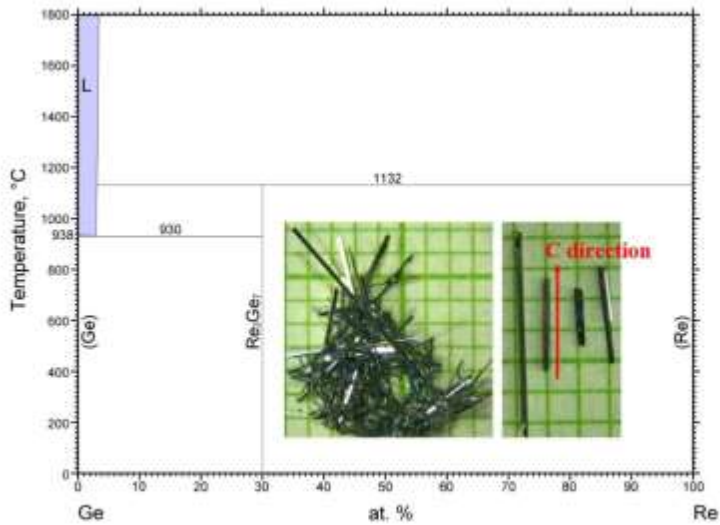

Figure 5: Binary phase diagram for Ge-Re with extremely sharp and narrow liquidus line for $Re_3Ge_7$ at Ge-rich concentrations. The inset shows single crystals of $Re_3Ge_7$ rods grown using CCS with the process outlined in the text.

**Harvesting of phase diagram data:  fuzzy or sharp.**

With increasing optimism and exuberance (irrational[33] or not) about machine learning and artificial intelligence (ML/AI) aiding in the prediction of complex, temperature-composition phase diagrams, there is a profound need (hunger even) for select data points on such diagrams to test, or anchor, such predictions.  Solution growth, in general, and CCS-based solution growth, *in particular*, can help provide exactly such data points.  Given that solution growth is premised upon interacting with a liquidus surface (or line) it can be very useful by providing information about the existence and/or location of this line.  As outlined above, we can use solution growth to delineate liquidus lines such as we did for the $GdCd_6$ and *i*-GdCd phases or identify misplaced liquidus lines, as was done for $La_2Ni_7$. Whereas these were specific binary phase diagrams that we had examined in detail, many solution growth experiments can provide sharp or "less-sharp" or even "fuzzy" data points about the ternary, quaternary, or higher phase diagram they are probing.  For example, when we were growing single crystals of $RAgSb_2$,[34] we were able to grow very large single crystals of $LaAgSb_2$ (Fig. 6 inset).  Going back to our lab-book entries we can find notes indicating that for an initial growth composition of $La_{0.05}Ag_{0.10}Sb_{0.85}$ (with initial masses of La: 0.6533 g; Ag: 1.0149 g; and Sb: 10.8187 g) we grew 1.62 g of single crystalline $LaAgSb_2$ by cooling from 1180 down to 750°C.  As a result, we can place a data point on the liquidus surface of $LaAgSb_2$ at $La_{0.015}Ag_{0.068}Sb_{0.917}$ at 750°C.  A second growth that decanted at 700°C with a similar large crystal allowed for a liquidus point to be inferred at $La_{0.012}Ag_{0.064}Sb_{0.924}$ at 700°C (Fig. 6).  If it were needed, higher temperature decants could help define the range of the liquidus surface intersected.  These data points come free of charge as the result of successful growth and decent record keeping.  In a similar manner, if a ternary or quaternary growth results in all of the material decanting into the catch side at a given temperature, then this means that a liquidus surface for primary solidification only might exists below the decanting temperature.  Finally, as long as there is only single phase growth, even if masses were not noted or could not be obtained, a less well defined composition that lies on the extrapolation of the line defined by the compound grown and the initial melt stoichiometry exists on the liquidus surface of the compound grown.  Depending on how well the mass of the single phase grown can be determined, or estimated, the composition along this line can be narrowed (made less fuzzy).   Looking over our decade's worth of 10,000's of growths, there are a lot of such data points that can be harvested and, as needed, refined or taken from fuzzy to sharper.

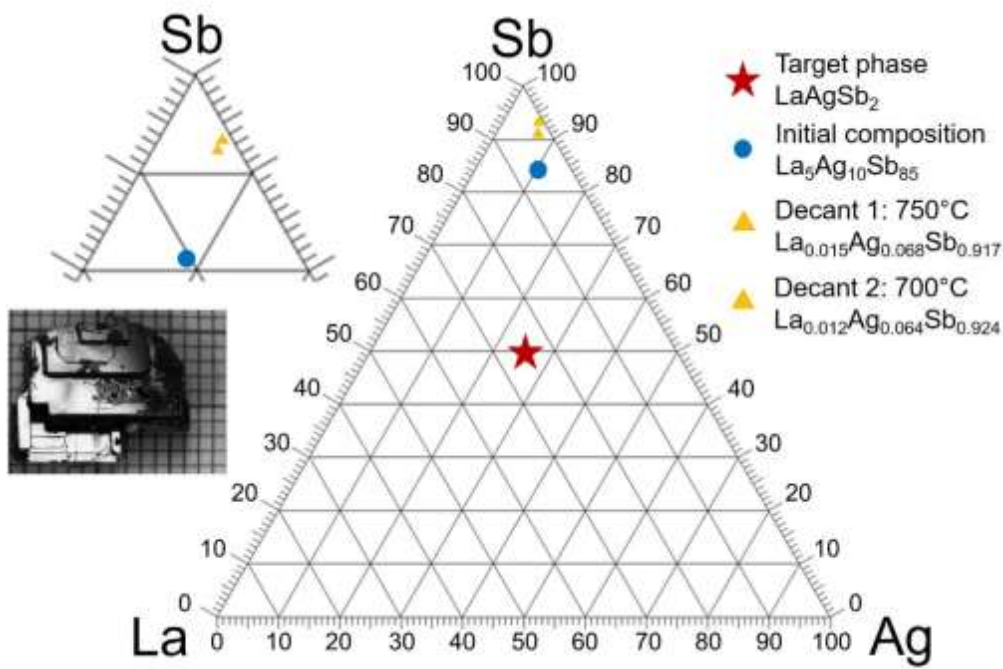

Figure 6: Ternary phase diagram for La-Ag-Sb. The red star marks the composition of the target phase, $LaAgSb_2$, the blue circle our initial starting composition, and the two gold triangles approximately mark two experimental (see text) points on the liquidus surface for crystallization of $LaAgSb_2$. The upper left inset shows a zoomed view of the points at Sb-rich compositions, and the lower inset is a large single crystal of $LaAgSb_2$[34] grown as discussed in the text.

With more common use of CCS for such growths, more recent growths offer less fuzzy data points than growths from over a decade ago (before the adoption of CCS as default for most growths).  At this point, as long as the growth results in a single phase solid, a data point on the multinary phase diagram can be created.  In addition, total spins (when there are no solids at the decanting temperature) provide upper limits to liquidus surfaces.  It would behoove humanity if these data points were made available for, _and used in_, AI/ML efforts to determine composition-temperature phase diagrams.  In addition, if, for example, there is a ternary phase diagram of particular interest, then data points such as these could be generated for the purpose of anchoring the computational phase diagram at a few select points.  Ideally, by doing so, such computationally based phase diagrams can be made more precise even for points well removed from the anchor points.  It is our hope, as experimentalists, that as great enthusiasm, and funding, is directed toward such computational efforts, the use of experimental data based on solution growth will be used to underpin and validate such efforts.

**Final notes and thoughts.**

Recently, LSP ceramics has been able to fabricate a modification of the CCS that is actually a throwback to the threaded frits that we had machined out of BN mentioned in Ref. 10.  Canfield - Svanidze Crucible Sets (CSCS) [35, 36] are a modification of the CCS that, in addition to the step-edge-frit-disc, also have threads on the crucible and the frit, which help to better contain the liquid during the decanting processes.  The somewhat courser threads and a different design allow for the production of the threaded CSCS in a manner that is still cheaper than two beer-company crucibles.  The threaded frit disk can, and does, still get glued in place in many cases, but generally, only to the growth side, not the catch side.  For cases when potential spillage from the growth side is a concern, the threaded CSCS systems offers a possible solution.

**Acknowledgements**

We want to thank Jiaqiang Yan for encouraging us to write this summary of CCS innovations up in a sooner is better than later manner.  This work was supported by the U.S. Department of Energy, Office of Basic Energy Science, Division of Materials Sciences and Engineering. The research was performed at the Ames Laboratory. Ames Laboratory is operated for the U.S. Department of Energy by Iowa State University under Contract No. DE-AC02-07CH11358.